\documentclass[twocolumn,showpacs,showkeys,prd,aps,amsmath,amssymb,10pt]{revtex4-1}
\usepackage{graphicx}
\usepackage{color}
\usepackage{multirow}
\usepackage{bm}
\usepackage{mathtools}
\usepackage[caption=false]{subfig}
\usepackage{url}
\usepackage{breakurl}
\usepackage{physics}
\usepackage{cancel}
\allowdisplaybreaks

\begin{document}
\title{Shell model study on the possibility of using an effective field theory for disentangling several contributions to the neutrinoless double-beta decay}
\author{Andrei Neacsu}
\email{neacs1a@cmich.edu}
\author{Mihai Horoi}
\email{mihai.horoi@cmich.edu}
\affiliation{Department of Physics, Central Michigan University, Mount Pleasant, Michigan 48859, USA}
\date{\today}
\begin{abstract}
Neutrinoless double-beta decay represents the most promising approach for revealing some of the most important, yet-unknown, properties of neutrinos related to their absolute masses and their nature. This transition involves beyond standard model theories that predict the violation of the lepton number conservation by two units. There is no experimental confirmation yet for this decay, but new experiments have set lower-limits for the associated half-lives in the case of several isotopes. 
Using an effective field theory that describes this transition, we calculate half-life ratios for five experimentally interesting isotopes in the case of 12 lepton number violating couplings.
These half-life ratios can be used to probe the sensitivity of the five isotopes in relation to their respective mechanisms, to predict the 
half-life limits needed to match the different experimental results, and in the case of experimental confirmation, these ratios could possibly indicate the dominant mechanism of the transition. 
We provide an analysis that could reveal valuable information regarding the dominant neutrinoless double-beta decay mechanism, if experimental half-life data becomes available for different isotopes.
\end{abstract}
\maketitle

\renewcommand*\arraystretch{1.3}
\section{Introduction}
The experimental discovery of neutrino oscillations~\cite{SuperKamiokande1998,SNO2001} was awarded the Nobel prize in 2015~\cite{kajita15,mcdonald15} for clarifying some of the properties of neutrinos. The important consequence of these observations is that neutrinos have non-zero mass. However, oscillation experiments alone can only measure squared mass differences, while other neutrino properties such as their mass hierarchy, their absolute masses, or their nature (whether neutrinos are Dirac or Majorana fermions) remain elusive. Nevertheless, the success of these experiments has greatly increased the interest in neutrino physics in general, and neutrinoless double-beta decay $(0\nu\beta\beta)$ in particular. The $0\nu\beta\beta$ is considered the best approach to further investigate the unknown properties mentioned. 
As such, large theoretical and experimental efforts are dedicated to the study of $0\nu\beta\beta$ transition, whcih requires the lepton number conservation be violated by two units. 

If $0\nu\beta\beta$ occurs, then the black-box theorems~\cite{SchechterValle1982,Nieves1984,Takasugi1984,Hirsch2006} can only state that the light left-handed neutrinos are Majorana particles, but do not provide a clear way to disentangle the possible contributions to this process. One of the most popular theories that takes into account the contribution of right-handed components in the beyond standard model (BSM) Lagrangian is the left-right symmetric model (LRSM)~\cite{PatiSalam1974,MohapatraPati1975,Senjanovic1975,KeungSenjanovic1983,Barry2013}, currently investigated in experiments at the Large Hadron Collider (LHC)~\cite{CMS2014} at CERN. 
Some contributions to the decay rate described by the LRSM could be disentangled and identified by measuring the angular and energy distributions of the outgoing electrons, and through the study of experimental half-life ratios of different isotopes~\cite{HoroiNeacsu2016prd,Neacsu2016ahep-dist}.
Because there could be other contributions that cannot be yet dismissed, a more general BSM effective field theory (EFT) would be desirable.

 Such an effective field theory was investigated and used in Ref.~\cite{Horoi2017eft} for the study of the neutrinoless double-beta decay. There, a thorough analysis of the LNV parameters was done using the latest experimental half-life limits of five nuclei under current investigation 
($T^{0\nu}_{^{48}\textmd{Ca}} > 2.0 \cdot 10^{22}$\cite{Nemo3-2016}, 
$T^{0\nu}_{^{76}\textmd{Ge}} > 8.0 \cdot 10^{25}$\cite{Gerda-2017-oct}, 
$T^{0\nu}_{^{82}\textmd{Se}} > 2.5 \cdot 10^{23}$\cite{Waters-2016}, 
$T^{0\nu}_{^{130}\textmd{Te}} > 4.0 \cdot 10^{24}$\cite{Cuore-2015}, and 
$T^{0\nu}_{^{136}\textmd{Xe}} > 1.07 \cdot 10^{26}$\cite{kamlandzen16}) and shell model nuclear matrix elements (NME)\cite{HoroiStoicaBrown2007,HoroiStoica2010,Senkov2016,NeacsuHoroi2016,NeacsuStoicaHoroi2012,
HoroiBrown2013,NeacsuStoica2014,SenkovHoroi2013,Horoi2013,BrownHoroiSenkov2014,SenkovHoroi2014,SenkovHoroiBrown2014,NeacsuHoroi2015}. 
With no exception, all the LNV parameter limits for $^{136}\textmd{Xe}$ were found to be the tightest ones. This feature was mainly because $^{136}\textmd{Xe}$ has the highest half-life lower-limit, which is larger then that of $^{76}\textmd{Ge}$, but also due to the interplay of corresponding NME and phase-space factors (PSF)~\cite{Stefanik2015,SuhonenCivitarese1998,Kotila2012,StoicaMirea2013,HoroiNeacsu2016psf}.

We further extend the use of the EFT of Ref.~\cite{Horoi2017eft} to calculate and study the half-life ratios for pairs from a number of five experimentally interesting isotopes in the case of 12 lepton number violating couplings. The ratio of half-lives can be used to probe the sensitivity of the five isotopes in relation to their respective mechanisms and to predict the half-life limits needed to match the different experimental results.
This information could be useful in estimating scales and costs, fine-tuning the experiments in search for the $0\nu\beta\beta$ transition mechanism that is expected to produce the shortest half-life, but also to get a better view and compare the status of various experiments.
Even more interesting is that in the case of experimental confirmation of $0\nu\beta\beta$ for different isotopes, one could possibly indicate the dominant mechanism of the transition.

In the present analysis we calculate the NME that enter the half-lives using shell model techniques and we take into account two sets of effective Hamiltonians and their corresponding optimal closure energies \cite{SenkovHoroi2014,SenkovHoroi2013,SenkovHoroiBrown2014}, $\left< E \right>$, specific for each model space. One set of NME is obtained using the Hamiltonians preferred by our group, and the results are designated by the "CMU" label. 
For $^{48}$Ca in the $pf$ model space $(0f_{7/2}, 1p_{3/2}, 0f_{5/2}, 1p_{1/2})$ we use GXPF1A~\cite{Honma2004} with $\left< E \right>=0.5$ MeV, for $^{76}$Ge and $^{82}$Se in the $jj44$ model space $(0f_{5/2}, 1p_{3/2}, 1p_{1/2}, 0g_{9/2})$ we choose JUN45~\cite{JUN45} with $\left< E \right>=3.4$ MeV, and for $^{130}$Te and $^{136}$Xe in the $jj55$ model space $(0g_{7/2}, 1d_{5/2}, 1d_{3/2}, 1s_{1/2}, 0h_{11/2})$ we use SVD~\cite{Chong2012} with $\left< E \right>=3.5$MeV. 
The second set of NME we calculate using the Hamiltonians preferred by the Strasbourg-Madrid group, and denoted with "St-Ma". 
 In this case, for $^{48}$Ca we use KB3G~\cite{KB3G} with $\left< E \right>=2.5$MeV, for $^{76}$Ge and $^{82}$Se GCN.28-50 with $\left< E \right>=10$MeV, and for $^{130}$Te and $^{136}$Xe we use GCN.50-82 with $\left< E \right>=12$MeV.
The choice of shell model NME is due to the fact that shell model calculations respect all the symmetries, take into account all the correlations around the Fermi surface, and could treat consistently the effects of the missing single-particle space using the many-body perturbation theory. The effects of the reduced model space were shown to be small, only about 20\% in the case of $^{82}$Se~\cite{HoltEngel2013}. Additionally, the effective Hamiltonians used in our shell model NME calculations have been tested by comparing several calculated observables with their experimental values (see Ref.~\cite{HoroiStoica2010, Senkov2016, NeacsuHoroi2016}) and by calculating the $2\nu\beta\beta$ NME that
were found to reproduce the experimental half-lives using a quenching factor of about 0.7~\cite{BrownFangHoroi2015}. No quenching of the bare operator was considered in the $0\nu\beta\beta$ NME calculations.

The PSF are calculated according to the effective method described in Ref.~\cite{HoroiNeacsu2016psf}. That method was proven to be very fast and reliable in reproducing the results of Ref.~\cite{Stefanik2015}, while allowing easy use of the electron kinematic factors used to calculate the electron angular and energy distributions.

This paper is organized as follows: Section~\ref{formalism} presents a very brief formalism of the $0\nu\beta\beta$ decay within the left-right symmetric model (LRSM), the $R$-parity violating SUSY model ($\cancel{\cal{R}}_p$), and the EFT. We present our results and discussion in Section~\ref{results}. The calculated half-lives that are expected to match the sensitivity of KamLand-Zen are shown in Subsection~\ref{halflives} and a possible way to extract the LNV parameters is presented in Subsection~\ref{ratios}. Lastly, Section~\ref{conclusions} is dedicated to the conclusions.

\section{Brief formalism of $0\nu\beta\beta$}\label{formalism}
The formalism described in this section was thoroughly reviewed in Ref.~\cite{Horoi2017eft}, but we briefly repeat the most important equations for clarity and for the convenience of the readers. For consistency, the labels and notations were kept identical to Ref.~\cite{Horoi2017eft}.

In the framework of the LRSM and $R$-parity violating SUSY model, the $0\nu\beta\beta$ half-life can be written as a sum of products 
of PSF, BSM LNV parameters, and their corresponding NME~\cite{HoroiNeacsu2016prd}:
\begin{flalign}\label{lrsm-hl}
\nonumber  \left[ T^{0\nu}_{1/2} \right] ^{-1}&= G_{01} g^4_A  \left| \eta_{0\nu}M_{0\nu} +\left(\eta^L_{N_R}+\eta^R_{N_R}\right)M_{0N} \right . \\
 & + \left . \eta_{\tilde{q}}M_{\tilde{q}} +\eta_{\lambda'}M_{\lambda'} +\eta_{\lambda}X_{\lambda}+\eta_{\eta}X_{\eta} \right| ^2  .
\end{flalign}
Here, $G_{01}$ is a phase space factor that can be calculated with good precision for most cases~\cite{Stefanik2015,SuhonenCivitarese1998,Kotila2012,StoicaMirea2013,HoroiNeacsu2016psf,HoroiNeacsu2016psf},
$g_A$ is the axial vector coupling constant, $\eta_{0\nu}=\frac{\left< m_{\beta\beta}\right>}{m_e}$ is the light left-handed neutrino parameter, with $\left< m_{\beta\beta}\right>$ representing the effective Majorana neutrino mass, and $m_e$ the electron mass. $\eta^L_{N_R}$, $\eta^R_{N_R}$ are the heavy neutrino parameters with left-handed and right-handed currents, respectively~\cite{Horoi2013,Barry2013}, 
$\eta_{\tilde{q}}$, $\eta_{\lambda'}$ are $\cancel{\cal{R}}_p$ SUSY LNV parameters~\cite{Vergados2012}, 
$\eta_{\lambda}$, and $\eta_{\eta}$ are parameters for the so-called ''$\lambda-$'' and ''$\eta-$mechanism'', respectively~\cite{Barry2013}. 
$M_{0\nu}$, $M_{0N}$, are the light and the heavy neutrino exchange NME, $M_{\tilde{q}}$, $M_{\lambda'}$ are the $\cancel{\cal{R}}_p$ SUSY NME, and
$X_{\lambda}$ and $X_{\eta}$ denote combinations of NME and other PSF ($G_{02}-G_{09}$) corresponding to the the $\lambda-$mechanism involving right-handed leptonic and right-handed hadronic currents, and the $\eta-$mechanism with right-handed leptonic and left-handed hadronic currents, respectively~\cite{HoroiNeacsu2016prd}.
Assuming a seesaw type I dominance \cite{DevMitra2015}, the term $\eta^L_{N_R}$ is considered not to contribute if the heavy mass eigenstates are larger than 1 GeV \cite{Blennow2010}, and we neglect it here. For consistency with the literature, the remaining term $\eta^R_{N_R}$ is labeled as $\eta_{0N}$.
 
A more general approach is based on the effective field theory extension of the Standard Model. 
The analysis based on the BSM contributions to the effective field theory is more desirable, because it does not rely on specific models, 
and their parameters could be extracted/constrained by the existing $0\nu\beta\beta$ data, and by data from LHC and other experiments. 
In fact, the models considered in Eq. (\ref{lrsm-hl}) always lead to a subset of terms in the low-energy ($\sim$ 200 MeV) effective field theory Lagrangian. 
Here we consider all the terms in the Lagrangian allowed by the symmetries. Some of the couplings will correspond to the model couplings of Eq. (\ref{lrsm-hl}), but they might have a wider meaning. Others are new, not corresponding to specific models.

In the case of the long-range component of the $0\nu\beta\beta$ diagram being treated as two point-like vertices at the Fermi scale exchanging a light neutrino, the dimension 6 Lagrangian can be expressed in terms of effective couplings~\cite{Deppisch2012}:
\begin{equation} \label{lag6}
 \mathcal{L}_6=\frac{G_F}{\sqrt{2}}\left[ j^\mu_{V-A} J^\dagger_{V-A,\mu} + \sum^{*}_{\alpha,\beta} \epsilon_\alpha^\beta j_\beta J^\dagger_\alpha \right] ,
\end{equation}
where $J^\dagger_\alpha=\bar{u}\mathcal{O}_\alpha d$ and $j_\beta=\bar{e}\mathcal{O}_\beta \nu$ are hadronic and leptonic Lorentz currents, respectively.
The definitions of the $\mathcal{O}_{\alpha,\beta}$ operators are given in Eq. (3) of Ref.~\cite{Deppisch2012}. The LNV parameters are 
$\epsilon_\alpha^\beta= \{ \epsilon^{V+A}_{V-A}, \ \epsilon^{V+A}_{V+A}, \ \epsilon^{S+P}_{S\pm P}, \ \epsilon^{TR}_{TR} \} $.
The ''*'' symbol indicates that the term with $\alpha=\beta=(V-A)$ is explicitly taken out of the sum. However, the first term in Eq. (\ref{lag6}) still entails BSM physics through the dimension-5 operator responsible for the Majorana neutrino mass.
Here $G_F=1.1663787\times 10^{-5}$ GeV$^{-2}$ denotes the Fermi coupling constant. 

In the short-range part of the diagram we consider the interaction to be point-like. 
Expressing the general Lorentz-invariant Lagrangian in terms of effective couplings~\cite{Hirsch1997}, we get:
\begin{flalign}\label{lag9}
\nonumber \mathcal{L}_9 & =\frac{G_F^2}{2 m_p} \biggl[ \varepsilon_1 JJj + \varepsilon_2 J^{\mu\nu} J_{\mu\nu}j + \varepsilon_3 J^{\mu} J_{\mu}j  \biggr .  \\
 & \biggl . + \varepsilon_4 J^{\mu} J_{\mu\nu}j^\nu + \varepsilon_5 J^{\mu} Jj_\mu \biggr] , 
\end{flalign}
with $m_p$ as the proton mass and the hadronic currents of defined chirality $J=\bar{u}(1\pm\gamma_5)d$, $J^\mu=\bar{u}\gamma^\mu(1\pm\gamma_5)d$, $J^{\mu\nu}=\bar{u}\frac{i}{2}[\gamma^\mu,\gamma^\nu](1\pm\gamma_5)d$, leptonic currents $j=\bar{e}(1\pm\gamma_5)e^C$, $j^\mu=\bar{e}\gamma^\mu(1\pm\gamma_5)e^C$, and $\varepsilon_\alpha^\beta=\varepsilon_\alpha^{xyz}=\{ \varepsilon_{1}, \ \varepsilon_{2}, \ \varepsilon_{3}^{LLz(RRz)}, \ \varepsilon_{3}^{LRz(RLz)}, \ \varepsilon_{4}, \ \varepsilon_{5} \}$.
These parameters have dependence on the chirality of the hadronic and the leptonic currents involved, with $xyz=L/R, L/R, L/R$.
In the case of $\varepsilon_3$, one can distinguish between different chiralities, thus we express them separately as
$\varepsilon_{3}^{LLz(RRz)}$ and $ \varepsilon_{3}^{LRz(RLz)}$.

The total number of LNV couplings in the EFT is 12: $\eta_{0\nu}$,  $\epsilon^{V+A}_{V-A}$, $\epsilon^{V+A}_{V+A}$, $\epsilon^{S+P}_{S\pm P}$, $\epsilon^{TR}_{TR}$, $\eta_{\pi\nu}$, $\varepsilon_{1}$, $\varepsilon_{2}$, $\varepsilon_{3}^{LLz(RRz)}$, $\varepsilon_{3}^{LRz(RLz)}$, $\varepsilon_{4}$, $\varepsilon_{5}$.

\section{Results and discussions} \label{results}

Under the assumption that a single mechanism/coupling dominates, we can reduce the expression of the half-life to the following form that allows us to extract limits and to put constraints on the LNV parameters:
\begin{equation}
 \left[ T^{0\nu}_{1/2}\right]^{-1}=g_{A}^4 \left| \eta_{LNV}\right|^2 {\cal{M}}_{LNV}^2 ,
\label{eq-onaxis}
\end{equation}
where $g_A=1.27$ is the axial-vector coupling constant, $\eta_{LNV}$ represents the effective LNV parameter, while the ${\cal{M}}_{LNV}^2$ coefficient contains the combination of NME and PFS associated to the LNV parameter. A full description of the ${\cal{M}}_{LNV}^2$ coefficients can be found in Ref. \cite{Horoi2017eft}
When using this equation to evaluate a single LNV parameter at a time from the experimental half-life limit, we call this an "on-axis analysis". Performing such a procedure is particularly useful in predicting the experimental sensitivity of different isotopes with respect to certain $0\nu\beta\beta$ 
mechanisms. 

The current best experimental $0\nu\beta\beta$ half-life limit is provided by the KamLand-Zen Collaboration $T^{1/2}_{^{136}\textmd{Xe}}=1.07 \cdot 10^{26}$~\cite{kamlandzen16}. Based on this limit, we extract on-axis the $\eta_{LNV}$ effective LNV parameters of interest for $^{136}$Xe.

Equation~(\ref{eq-onaxis}) is factorized in such a way that the half-life ratio of two isotopes, or even between different mechanisms assumed to dominate the transition in the same isotope, can be easily obtained simply from the inverse of the ${\cal{M}}_{LNV}^2$ ratio. This eliminates the need to rely on any experimental half-life limit, or extracted LNV parameters for evaluating the relative sensitivity of different isotopes to various mechanisms. 
We present our analysis and results from two perspectives: on one side, we evaluate the equivalent half-lives required to match the current $^{136}$Xe sensitivity to the LNV parameters, and on the other side, we search for the best half-life ratios that are expected to help distinguish between the different possible contributions to the decay rate, under the one mechanism/coupling dominance assumption.

We calculate the NME components of the ${\cal{M}}_{LNV}^2$ coefficients within the interacting shell model (ISM). For each of the isotopes we obtain results using two sets of effective Hamiltonians, denoted as "CMU" and "St-Ma", which are popular in the literature and have been thoroughly tested against experimental data. Their description is given in the Introduction.
All the NME used in this analysis were calculated using a Jastrow short-range correlation (SRC) method using the CD-Bonn and AV-18 parametrizations (see e.g. Ref. \cite{HoroiStoica2010} and Refs. therein).
Different choices for the method of including the SRC affect the values of the NME, as has been discussed in many papers in recent literature.
One can find a graphical representation of these effects on the NME in Fig. 6 and Fig. 7 of Ref.~\cite{NeacsuHoroi2016} for the light left-handed neutrino exchange and for the heavy right-handed neutrino exchange, respectively. Although the differences in NME values can be large, in the case of shell model calculations, the ratio of ${\cal{M}}_{LNV}^2$ coefficients are usually stable when calculated consistently with the same parametrization. This behavior is detailed in the following subsections.

The PSF components are calculated using a recently proposed effective method described in great detail in Ref.~\cite{HoroiNeacsu2016psf}. $G_{01}$ values were calculated with a screening factor ($s_f$) of 94.5, while for $G_{02}-G_{09}$ we used $s_f=92.0$ that was shown to provide results very close to those of Ref.~\cite{Stefanik2015}.

\subsection{Calculated half-lives for $^{48}$Ca, $^{76}$Ge, $^{82}$Se, and $^{130}$Te expected to match the current $^{136}$Xe sensitivity}\label{halflives}
In this subsection we extract the LNV parameters from the current $^{136}$Xe experimental limit of $1.07 \cdot 10^{26}$ years \cite{kamlandzen16}. We take into account two SRC parameterizations (CD-Bonn and AV-18) and two sets of effective Hamiltonians. The values are shown in the upper parts of all the tables presented in this subsection as $\eta_{LNV}$. We use these parameters together with shell model NME to predict the half-life limits of $^{48}$Ca, $^{76}$Ge, $^{82}$Se, and $^{130}$Te that are needed to match the current limit for $^{136}$Xe for each of the LNV mechanisms. This could prove particularly useful in determining the minimum required isotope quantities needed by the experiments in order to obtain better constraints on the LNV parameters. This information will, however, need to be adjusted to the particular setup of each experiment and correlated with their detection efficiency of $0\nu\beta\beta$ transitions for their isotopes of choice.

For the results presented in the tables of this Section, we preserve the notations of Ref.~\cite{Horoi2017eft}. As such, the LNV parameters corresponding to  LRSM and $\cancel{\cal{R}}_p$ are denoted as $\eta_\alpha$, the ones corresponding to the dimension 6 Lagrangian are $\epsilon_\alpha^\beta$, and, lastly, the ones coming from the dimension 9 Lagrangian are written as $\varepsilon_\alpha^\beta$. This choice eliminates the need for repeating the complete formalism, diminishes the risk of confusion related to changes of notation, and allows for a more accessible read of the underlying details presented in that reference.

Table~\ref{tab-models} presents upper-limit values of the $\eta_\alpha$ LNV parameters for the LRSM and $\cancel{\cal{R}}_p$ extracted on-axis from the most recent $^{136}$Xe experiment~\cite{kamlandzen16}, the CMU ${\cal{M}}_{LNV}^2$ coefficients containing NME and PSF, and the calculated half-life limits for the other isotopes of experimental interest predicted when taking into account the $^{136}$Xe $\eta_\alpha$ LNV parameters. The effect of the SRC parameterizations on the different NME can be easily seen in this table. The ${\cal{M}}_{LNV}^2$ coefficients can almost double when changing from AV-18 to CD-Bonn in the case of ${\cal{M}}^2_{0N}$. Nevertheless, the predicted half-life limits are very stable if the extracted $^{136}$Xe LNV parameters are chosen from calculations with the same SRC.
\begin{table*}[htb] 
    \centering
 \begin{minipage}{0.70\linewidth}
 \caption{The upper part presents the $\eta_\alpha$ LNV parameters for the LRSM and SUSY extracted on-axis from the most recent $^{136}$Xe experiment~\cite{kamlandzen16}, the middle part shows our shell model ${\cal{M}}_{LNV}^2$ coefficients containing NME and PSF~\cite{Horoi2017eft}, and the lower part lists the calculated half-life limits in years for the isotopes of experimental interest predicted when taking into account the $^{136}$Xe LNV parameters. The NME are calculated with the CMU effective Hamiltonians and two SRC parameterizations: CD-Bonn and AV-18.}
 \begin{tabular}{lccccccc} \label{tab-models}\smallskip
SRC &$ \eta_{LNV}$&$ |\eta_{0\nu}|\cdot 10^{7}$&$|\eta_{0N}|\cdot 10^{9}$&$|\eta_{\lambda}|\cdot 10^{7}$&$|\eta_{\eta}|\cdot 10^{9}$&$|\eta_{\tilde{q}}|\cdot 10^{9}$&$|\eta_{\lambda^\prime}|\cdot 10^{9}$ 	\\ 
CD-Bonn &  &  2.86 &  4.40 &  2.01 &  1.08 &  2.95 &  2.24 \\ 
AV-18   &  &  3.08 &  6.11 &  2.17 &  1.31 &  3.10 &  2.12 \\  \hline \smallskip
& ${\cal{M}}_{LNV}^2$ &${\cal{M}}^2_{0\nu}\cdot 10^{14}$&${\cal{M}}^2_{0N}\cdot 10^{10}$&${\cal{M}}^2_{\lambda}\cdot 10^{13}$&${\cal{M}}^2_{\eta}\cdot 10^{9}$&${\cal{M}}^2_{\tilde{q}}\cdot 10^{10}$&${\cal{M}}^2_{\lambda^\prime}\cdot 10^{10}$ 	\\ 
\multirow{5}{*}{CD-Bonn} 
 & $^{48}$Ca   &  2.57 &  1.63 &  1.09 &  1.45 &  2.83 &  6.44 \\ 
 & $^{76}$Ge   &  3.00 &  0.87 &  0.39 &  1.40 &  2.62 &  3.99 \\ 
 & $^{82}$Se   &  11.5 &  3.28 &  3.21 &  5.11 &  10.20&  15.6 \\ 
 & $^{130}$Te  &  5.22 &  2.25 &  1.11 &  3.67 &  4.85 &  8.43 \\  \smallskip
 & $^{136}$Xe  &  4.40 &  1.86 &  0.89 &  3.09 &  4.13 &  7.14 \\
\multirow{5}{*}{AV-18} 
 & $^{48}$Ca   &  2.19 &  0.94 &  0.90 &  0.92 &  2.57 &  7.08 \\ 
 & $^{76}$Ge   &  2.67 &  0.46 &  0.34 &  0.95 &  2.41 &  4.40 \\ 
 & $^{82}$Se   &  10.3 &  1.75 &  2.85 &  3.47 &  9.43 &  17.1 \\ 
 & $^{130}$Te  &  4.49 &  1.17 &  0.96 &  2.48 &  4.38 &  9.43 \\ 
 & $^{136}$Xe  &  3.79 &  0.96 &  0.77 &  2.10 &  3.73 &  7.97 \\   \hline \smallskip
& $T_{LNV}\cdot 10^{-26}$&$T_{0\nu}$&$T_{0N}$&$T_{\lambda}$&$T_{\eta}$&$T_{\tilde{q}}$&$T_{\lambda^\prime}$ \\ 
\multirow{4}{*}{CD-Bonn}
 & $^{48}$Ca   &  1.83 &  1.22 &  0.87 &  2.28 &  1.56 &  1.19 \\ 
 & $^{76}$Ge   &  1.57 &  2.28 &  2.45 &  2.36 &  1.69 &  1.92 \\ 
 & $^{82}$Se   &  0.41 &  0.61 &  0.30 &  0.65 &  0.43 &  0.49 \\  \smallskip
 & $^{130}$Te  &  0.90 &  0.88 &  0.85 &  0.90 &  0.91 &  0.91 \\
\multirow{4}{*}{AV-18}  
 & $^{48}$Ca   &  1.85 &  1.09 &  0.91 &  2.45 &  1.55 &  1.20 \\ 
 & $^{76}$Ge   &  1.52 &  2.22 &  2.39 &  2.37 &  1.66 &  1.94 \\ 
 & $^{82}$Se   &  0.39 &  0.59 &  0.29 &  0.65 &  0.42 &  0.50 \\ 
 & $^{130}$Te  &  0.90 &  0.88 &  0.86 &  0.90 &  0.91 &  0.90 \\   \hline
 \end{tabular}
 \end {minipage}
 \end{table*}
 
Using the St-Ma effective Hamiltonians, in Table~\ref{tab-models-2} we recalculate the quantities from Table~\ref{tab-models}. The same observations related to the SRC are valid also in this case. The most significant ${\cal{M}}_{LNV}^2$ change occurs for the ${\cal{M}}^2_{0N}$ coefficient, but the predicted half-life limits remain stable. Regarding how the effective Hamiltonians affect the NME, one can see that KB3G preferred by St-Ma provides higher values for $^{48}$Ca than GXPF1A prefferred by us (CMU), GCN.28-50 yields lower NME than JUN45 for $^{76}$Ge and $^{82}$Se, and GCN.50:82 results in higher NME than SVD for $^{130}$Te and $^{136}$Xe. 
\begin{table*}[htb] 
    \centering
 \begin{minipage}{0.70\linewidth}
 \caption{
Same as Table~\ref{tab-models}, but with ${\cal{M}}_{LNV}^2$ coefficients obtained using St-Ma Hamiltonians.}
 \begin{tabular}{llcccccc} \label{tab-models-2} \smallskip
SRC&$ \eta_{LNV}$&$ |\eta_{0\nu}|\cdot 10^{-7}$&$|\eta_{0N}|\cdot 10^{-9}$&$|\eta_{\lambda}|\cdot 10^{-7}$&$|\eta_{\eta}|\cdot 10^{-9}$&$|\eta_{\tilde{q}}|\cdot 10^{-9}$&$|\eta_{\lambda^\prime}|\cdot 10^{-9}$ 	\\ 
CD-Bonn &  &  2.09 &  3.14 &  1.62 &  0.82 &  2.19 &  1.61 \\ 
AV-18   &  &  2.24 &  4.34 &  1.74 &  1.00 &  2.30 &  1.52 \\ 	\hline \smallskip
& ${\cal{M}}_{LNV}^2$ &${\cal{M}}^2_{0\nu}\cdot 10^{14}$&${\cal{M}}^2_{0N}\cdot 10^{10}$&${\cal{M}}^2_{\lambda}\cdot 10^{13}$&${\cal{M}}^2_{\eta}\cdot 10^{9}$&${\cal{M}}^2_{\tilde{q}}\cdot 10^{10}$&${\cal{M}}^2_{\lambda^\prime}\cdot 10^{10}$ 	\\ 
\multirow{5}{*}{CD-Bonn} 
 & $^{48}$Ca   &  3.13 &  1.91 &  1.48 &  1.80 &  3.19 &  7.43 \\ 
 & $^{76}$Ge   &  1.98 &  0.71 &  0.22 &  1.05 &  1.73 &  3.01 \\ 
 & $^{82}$Se   &  7.54 &  2.57 &  1.84 &  3.74 &  6.51 &  11.30 \\ 
 & $^{130}$Te  &  12.5 &  5.57 &  2.19 &  8.29 &  11.4 &  21.3 \\ \smallskip
 & $^{136}$Xe  &  8.22 &  3.64 &  1.38 &  5.35 &  7.47 &  13.9 \\ 
 \multirow{5}{*}{AV-18} 
 & $^{48}$Ca   &  2.66 &  1.08 &  1.23 &  1.13 &  2.87 &  8.21 \\ 
 & $^{76}$Ge   &  1.75 &  0.38 &  0.19 &  0.70 &  1.58 &  3.33 \\ 
 & $^{82}$Se   &  6.68 &  1.37 &  1.62 &  2.52 &  5.97 &  12.5 \\ 
 & $^{130}$Te  &  10.8 &  2.92 &  1.89 &  5.58 &  10.3 &  23.8 \\ 
 & $^{136}$Xe  &  7.14 &  1.91 &  1.19 &  3.60 &  6.78 &  15.5 \\ \hline \smallskip
&$T_{LNV}\cdot 10^{-26}$	&$T_{0\nu}$&$T_{0N}$	&$T_{\lambda}$	&$T_{\eta}$	&$T_{\tilde{q}}$&$T_{\lambda^\prime}$ \\ 
\multirow{4}{*}{CD-Bonn}
 & $^{48}$Ca   &  2.81 &  2.04 &  1.00 &  3.17 &  2.51 &  2.01 \\ 
 & $^{76}$Ge   &  4.44 &  5.49 &  6.65 &  5.43 &  4.62 &  4.96 \\ 
 & $^{82}$Se   &  1.17 &  1.52 &  0.80 &  1.53 &  1.23 &  1.32 \\ \smallskip
 & $^{130}$Te  &  0.70 &  0.70 &  0.67 &  0.69 &  0.70 &  0.70 \\ 
 \multirow{4}{*}{AV-18}
 & $^{48}$Ca   &  2.87 &  1.90 &  1.03 &  3.41 &  2.52 &  2.03 \\ 
 & $^{76}$Ge   &  4.37 &  5.41 &  6.57 &  5.47 &  4.57 &  4.99 \\ 
 & $^{82}$Se   &  1.14 &  1.49 &  0.79 &  1.53 &  1.21 &  1.33 \\ 
 & $^{130}$Te  &  0.71 &  0.70 &  0.67 &  0.69 &  0.70 &  0.70 \\  \hline
 \end{tabular}
 \end {minipage}
 \end{table*}

 Within the framework of the EFT, in Table~\ref{tab-longrange} we investigate the ${\cal{M}}_{LNV}^2$ coefficients, and the half-lives corresponding to the dimension 6 Lagrangian in Eq.~(\ref{lag6}). 
In this case, the $|\epsilon^{V+A}_{V-A}|$ LNV parameters, the ${\cal{M}}^2_{{V+A}/{V-A}}$ coefficients, and the $T_{{V+A}/{V-A}}$ half-lives of the EFT also correspond to the so-called "$\eta-$mechanism" in the LRSM and presented in Table~\ref{tab-models}. Similarly the $|\epsilon^{V+A}_{V+A}|$ LNV parameters, the ${\cal{M}}^2_{{V+A}/{V+A}}$ coefficients, the $T_{{V+A}/{V+A}}$ half-lives, correspond the so-called "$\lambda-$mechanism". In Ref.~\cite{Horoi2017eft} it was shown that one can obtain another alternative value for $|\epsilon^{TR}_{TR}|$, $|\tilde{\epsilon}_{TR}^{TR}|=|\eta_{\pi\nu}| /8$, where our $\eta_{\pi\nu}$ plays the same role as $\eta^{11}_{(q)LR}$ in Eq.(22) of Ref.~\cite{Kovalenko2008} and $\eta_{\bar{q}}$ in Table~\ref{tab-models} (see also Eq. (154) of Ref.~\cite{Vergados2012}). Here, we notice a significant effect of the SRC choice on the NME, but also on the predicted half-life limits, especially in the case of $^{48}$Ca. The $^{136}$Xe alternative $|\tilde{\epsilon}_{TR}^{TR}|=|\eta_{\pi\nu}| /8$ LNV parameter is similar to $|\epsilon_{TR}^{TR}|$, but the associated NME and half-life limits are very stable with respect to the choice of SRC.
\begin{table*}
 \caption{Same as Table~\ref{tab-models}, but for the long-range contribution to the $0\nu\beta\beta$ diagram, corresponding to the dimension 6 Lagrangian for the CMU set of Hamiltonians.}
 \begin{tabular}{llcccccc}  \label{tab-longrange}\smallskip
SRC&$ \eta_{LNV}$	&$|\epsilon^{V+A}_{V-A}|\cdot 10^{9}$&$|\epsilon^{V+A}_{V+A}|\cdot 10^{7}$&$|\epsilon^{S+P}_{S\pm P}|\cdot 10^{9}$&$|\epsilon^{TR}_{TR}|\cdot 10^{10}$&$|\eta_{\pi\nu}| \cdot 10^{9}$ 	\\ 
CD-Bonn  &  &  1.08 &  2.01 &  4.09 &  3.59 &  2.95 \\ 
AV-18    &  &  1.31 &  2.17 &  5.65 &  4.63 &  3.10 \\   \hline \smallskip
&${\cal{M}}^2_{LNV}$	&${\cal{M}}^2_{{V+A}/{V-A}}\cdot 10^{14}$&${\cal{M}}^2_{{V+A}/{V+A}}\cdot 10^{10}$&${\cal{M}}^2_{{S+P}/{S\pm P}}\cdot 10^{10}$&${\cal{M}}^2_{{TR}/{TR}}\cdot 10^{8}$&${\cal{M}}^2_{\pi\nu}\cdot 10^{10}$ 	\\
\multirow{5}{*}{CD-Bonn} 
 & $^{48}$Ca   &  1.45 &  1.09 &  8.85 &  0.25 &  2.83 \\ 
 & $^{76}$Ge   &  1.40 &  0.39 &  1.33 &  1.09 &  2.62 \\ 
 & $^{82}$Se   &  5.11 &  3.21 &  5.12 &  4.04 &  10.2 \\ 
 & $^{130}$Te  &  3.67 &  1.11 &  2.72 &  3.28 &  4.85 \\ \smallskip
 & $^{136}$Xe  &  3.09 &  0.89 &  2.15 &  2.79 &  4.13 \\ 
  \multirow{5}{*}{AV-18} 
 & $^{48}$Ca   &  0.92 &  0.90 &  7.03 &  0.04 &  2.57 \\ 
 & $^{76}$Ge   &  0.95 &  0.34 &  0.78 &  0.63 &  2.41 \\ 
 & $^{82}$Se   &  3.47 &  2.85 &  3.05 &  2.34 &  9.43 \\ 
 & $^{130}$Te  &  2.48 &  0.96 &  1.45 &  1.96 &  4.38 \\ 
 & $^{136}$Xe  &  2.10 &  0.77 &  1.13 &  1.67 &  3.73 \\  \hline \smallskip
&$T_{LNV}\cdot 10^{-26}$	&$T_{{V+A}/{V-A}}$&$T_{{V+A}/{V+A}}$&$T_{{S+P}/{S\pm P}}$&$T_{{TR}/{TR}}$&$T_{\pi\nu}$ \\ 
\multirow{4}{*}{CD-Bonn} 
 & $^{48}$Ca   &  2.28 &  0.87 &  0.26 &  12.0 &  1.56 \\ 
 & $^{76}$Ge   &  2.36 &  2.45 &  1.73 &  2.74 &  1.69 \\ 
 & $^{82}$Se   &  0.65 &  0.30 &  0.45 &  0.74 &  0.43 \\ \smallskip
 & $^{130}$Te  &  0.90 &  0.85 &  0.85 &  0.91 &  0.91 \\ 
 \multirow{4}{*}{AV-18} 
 & $^{48}$Ca   &  2.45 &  0.91 &  0.17 &  45.1 &  1.55 \\ 
 & $^{76}$Ge   &  2.37 &  2.39 &  1.54 &  2.83 &  1.66 \\ 
 & $^{82}$Se   &  0.65 &  0.29 &  0.40 &  0.76 &  0.42 \\ 
 & $^{130}$Te  &  0.90 &  0.86 &  0.83 &  0.91 &  0.91 \\   \hline
 \end{tabular}
 \end{table*} 
 
For the second set of Hamiltonians, we present the results corresponding to the dimension 6 Lagrangian of Eq.~(\ref{lag6}) in Table~\ref{tab-longrange-2}. The same conclusions and observations that we made for Table~\ref{tab-longrange} are also valid here.

\begin{table*}
 \caption{Same as Table~\ref{tab-longrange}, but with ${\cal{M}}_{LNV}^2$ coefficients calculated with St-Ma Hamiltonians.}
 \begin{tabular}{llcccccc}  \label{tab-longrange-2}\smallskip
SRC&$ \eta_{LNV}$	&$|\epsilon^{V+A}_{V-A}|\cdot 10^{9}$&$|\epsilon^{V+A}_{V+A}|\cdot 10^{7}$&$|\epsilon^{S+P}_{S\pm P}|\cdot 10^{9}$&$|\epsilon^{TR}_{TR}|\cdot 10^{10}$&$|\eta_{\pi\nu}| \cdot 10^{9}$ 	\\ 
CD-Bonn   &  &  0.82 &  1.62 &  2.86 &  2.80 &  2.19 \\ 
AV-18	  &  &  1.00 &  1.74 &  3.87 &  3.67 &  2.30 \\    \hline \smallskip
&${\cal{M}}^2_{LNV}$	&${\cal{M}}^2_{{V+A}/{V-A}}\cdot 10^{14}$&${\cal{M}}^2_{{V+A}/{V+A}}\cdot 10^{10}$&${\cal{M}}^2_{{S+P}/{S\pm P}}\cdot 10^{10}$&${\cal{M}}^2_{{TR}/{TR}}\cdot 10^{8}$&${\cal{M}}^2_{\pi\nu}\cdot 10^{10}$ 	\\
\multirow{5}{*}{CD-Bonn} 
 & $^{48}$Ca   &  1.80 &  1.48 &  7.50 &  0.73 &  3.19 \\ 
 & $^{76}$Ge   &  1.05 &  0.22 &  1.03 &  0.81 &  1.73 \\ 
 & $^{82}$Se   &  3.74 &  1.84 &  3.63 &  3.01 &  6.51 \\ 
 & $^{130}$Te  &  8.29 &  2.19 &  6.65 &  7.09 &  11.40 \\ \smallskip
 & $^{136}$Xe  &  5.35 &  1.38 &  4.39 &  4.58 &  7.47 \\ 
  \multirow{5}{*}{AV-18} 
 & $^{48}$Ca   &  1.13 &  1.23 &  5.63 &  0.26 &  2.87 \\ 
 & $^{76}$Ge   &  0.70 &  0.19 &  0.61 &  0.46 &  1.58 \\ 
 & $^{82}$Se   &  2.52 &  1.62 &  2.11 &  1.73 &  5.97 \\ 
 & $^{130}$Te  &  5.58 &  1.89 &  3.61 &  4.13 &  10.30 \\ 
 & $^{136}$Xe  &  3.60 &  1.19 &  2.40 &  2.66 &  6.78 \\   \hline \smallskip
&$T_{LNV}\cdot 10^{-26}$	&$T_{{V+A}/{V-A}}$&$T_{{V+A}/{V+A}}$&$T_{{S+P}/{S\pm P}}$&$T_{{TR}/{TR}}$&$T_{\pi\nu}$ \\ 
\multirow{4}{*}{CD-Bonn} 
 & $^{48}$Ca   &  3.17 &  1.00 &  0.63 &  6.74 &  2.51 \\ 
 & $^{76}$Ge   &  5.43 &  6.65 &  4.55 &  6.03 &  4.62 \\ 
 & $^{82}$Se   &  1.53 &  0.80 &  1.29 &  1.63 &  1.23 \\ \smallskip
 & $^{130}$Te  &  0.69 &  0.67 &  0.71 &  0.69 &  0.70 \\ 
 \multirow{4}{*}{AV-18} 
 & $^{48}$Ca   &  3.41 &  1.03 &  0.46 &  10.80 &  2.52 \\ 
 & $^{76}$Ge   &  5.47 &  6.57 &  4.23 &  6.15 &  4.57 \\ 
 & $^{82}$Se   &  1.53 &  0.79 &  1.21 &  1.64 &  1.21 \\ 
 & $^{130}$Te  &  0.69 &  0.67 &  0.71 &  0.69 &  0.70 \\    \hline
 \end{tabular}
 \end{table*} 
 
The results corresponding to the dimension 9 Lagrangian of Eq.~(\ref{lag9}) are displayed in Table~\ref{tab-shortrange} as in the previous tables. Regarding $|\varepsilon_{3}^{RRz(LLz)}|$, we note that the results closely correspond to the ones for $|\eta_{0N}|$ in Table~\ref{tab-models} for the LRSM, but the tensor component of the NME is missing in the formalism associated to this case. 
The NME that enter the ${\cal{M}}^2_{LNV}$ coefficients manifest the same behavior as the the ${\cal{M}}^2_{0N}$ when changing SRC parameterizations, but the predicted half-life limits are very stable.
 \begin{table*}
  \caption{Same as Tables~\ref{tab-models} and~\ref{tab-longrange}, for the short-range contribution to the $0\nu\beta\beta$ diagram, corresponding to the dimension 9 Lagrangian for the CMU set of Hamiltonians.}
 \begin{tabular}{llcccccccc} \label{tab-shortrange}   \smallskip
SRC&$ \eta_{LNV}$	&$|\varepsilon_{1}|\cdot 10^{8}$&$|\varepsilon_{2}|\cdot 10^{10}$&$|\varepsilon_{3}^{RRz(LLz)}|\cdot 10^{9}$& $|\varepsilon_{3}^{LRz(RLz)}|\cdot 10^{9}$&$|\varepsilon_{4}|\cdot 10^{9}$&$|\varepsilon_{5}|\cdot 10^{8}$ &$|\eta_{\pi N}| \cdot 10^{9}$	\\ 
CD-Bonn &  &  9.44 &  5.70 &  4.31 &  7.15 &  5.00 &  4.58 &  2.24 \\ 
AV-18   &  &  11.4 &  8.14 &  5.95 &  10.9 &  7.15 &  5.52 &  2.12 \\  \hline  \smallskip
&${\cal{M}}^2_{LNV}$&${\cal{M}}^2_{1}\cdot 10^{13}$&${\cal{M}}^2_{2}\cdot 10^{8}$&${\cal{M}}^2_{3/LLz(RRz)}\cdot 10^{10}$&${\cal{M}}^2_{3/LRz(RLz)}\cdot 10^{11}$&${\cal{M}}^2_{4}\cdot 10^{10}$&${\cal{M}}^2_{5}\cdot 10^{12}$&${\cal{M}}^2_{\pi N}\cdot 10^{10}$ 	\\
\multirow{5}{*}{CD-Bonn} 
 & $^{48}$Ca   &  2.63 &  0.68 &  1.21 &  4.27 &  0.91 &  1.15 &  6.44 \\ 
 & $^{76}$Ge   &  1.83 &  0.50 &  0.87 &  3.16 &  0.70 &  0.84 &  3.99 \\ 
 & $^{82}$Se   &  6.86 &  1.87 &  3.26 &  11.8 &  2.50 &  3.01 &  15.6 \\ 
 & $^{130}$Te  &  4.83 &  1.34 &  2.33 &  8.50 &  1.74 &  2.06 &  8.43 \\ \smallskip
 & $^{136}$Xe  &  4.03 &  1.11 &  1.93 &  7.03 &  1.44 &  1.71 &  7.14 \\  
\multirow{5}{*}{AV-18}
 & $^{48}$Ca   &  1.81 &  0.33 &  0.63 &  1.81 &  0.44 &  0.80 &  7.08 \\ 
 & $^{76}$Ge   &  1.27 &  0.25 &  0.46 &  1.38 &  0.35 &  0.58 &  4.40 \\ 
 & $^{82}$Se   &  4.77 &  0.93 &  1.74 &  5.19 &  1.24 &  2.09 &  17.1 \\ 
 & $^{130}$Te  &  3.32 &  0.65 &  1.22 &  3.66 &  0.85 &  1.41 &  9.43 \\ 
 & $^{136}$Xe  &  2.77 &  0.54 &  1.02 &  3.03 &  0.70 &  1.18 &  7.97 \\ \hline \smallskip
&$T_{LNV}\cdot 10^{-26}$	&$T_{1}$&$T_{2}$	&$T_{3/RRz(LLz)}$&$T_{3/LRz(RLz)}$&$T_{4}$	&$T_{5}$&$T_{\pi N}$ \\ 
\multirow{4}{*}{CD-Bonn} 
& $^{48}$Ca   &  1.64 &  1.73 &  1.71 &  1.76 &  1.68 &  1.59 &  1.19 \\ 
 & $^{76}$Ge   &  2.36 &  2.38 &  2.38 &  2.38 &  2.20 &  2.18 &  1.92 \\ 
 & $^{82}$Se   &  0.63 &  0.63 &  0.63 &  0.64 &  0.62 &  0.61 &  0.49 \\ \smallskip
 & $^{130}$Te  &  0.89 &  0.89 &  0.89 &  0.89 &  0.88 &  0.89 &  0.91 \\ 
\multirow{4}{*}{AV-18}
 & $^{48}$Ca   &  1.64 &  1.74 &  1.72 &  1.79 &  1.69 &  1.59 &  1.20 \\ 
 & $^{76}$Ge   &  2.35 &  2.35 &  2.35 &  2.35 &  2.17 &  2.17 &  1.94 \\ 
 & $^{82}$Se   &  0.62 &  0.62 &  0.62 &  0.63 &  0.61 &  0.60 &  0.50 \\ 
 & $^{130}$Te  &  0.89 &  0.89 &  0.89 &  0.89 &  0.89 &  0.89 &  0.90 \\  \hline
 \end{tabular}
 \end{table*}

 Finally, we show our results corresponding to the dimension 9 Lagrangian of Eq.~(\ref{lag9}) when using the second set of Hamiltonians in Table~\ref{tab-shortrange-2}. The dependence of the ${\cal{M}}^2_{LNV}$ coefficients and of the predicted half-life limits is similar to that found in Table~\ref{tab-shortrange}.
 \begin{table*}
  \caption{Same as Table~\ref{tab-shortrange}, for the short-range contribution to the $0\nu\beta\beta$ diagram, corresponding to the dimension 9 Lagrangian for the St-Ma set of Hamiltonians.}
 \begin{tabular}{llcccccccc} \label{tab-shortrange-2}   \smallskip
SRC&$ \eta_{LNV}$	&$|\varepsilon_{1}|\cdot 10^{8}$&$|\varepsilon_{2}|\cdot 10^{10}$&$|\varepsilon_{3}^{RRz(LLz)}|\cdot 10^{9}$& $|\varepsilon_{3}^{LRz(RLz)}|\cdot 10^{9}$&$|\varepsilon_{4}|\cdot 10^{9}$&$|\varepsilon_{5}|\cdot 10^{8}$ &$|\eta_{\pi N}| \cdot 10^{9}$	\\ 
CD-Bonn  &  &  6.77 &  4.12 &  3.12 &  5.19 &  3.62 &  3.28 &  1.61 \\ 
AV-18    &  &  8.15 &  5.89 &  4.29 &  7.90 &  5.17 &  3.95 &  1.52 \\   \hline  \smallskip
&${\cal{M}}^2_{LNV}$&${\cal{M}}^2_{1}\cdot 10^{13}$&${\cal{M}}^2_{2}\cdot 10^{8}$&${\cal{M}}^2_{3/LLz(RRz)}\cdot 10^{10}$&${\cal{M}}^2_{3/LRz(RLz)}\cdot 10^{11}$&${\cal{M}}^2_{4}\cdot 10^{10}$&${\cal{M}}^2_{5}\cdot 10^{12}$&${\cal{M}}^2_{\pi N}\cdot 10^{10}$ 	\\
\multirow{5}{*}{CD-Bonn} 
 & $^{48}$Ca   &  3.31 &  0.89 &  1.56 &  5.61 &  1.19 &  1.45 &  7.43 \\ 
 & $^{76}$Ge   &  1.47 &  0.40 &  0.70 &  2.55 &  0.57 &  0.68 &  3.01 \\ 
 & $^{82}$Se   &  5.38 &  1.46 &  2.56 &  9.26 &  1.96 &  2.36 &  11.3 \\ 
 & $^{130}$Te  &  12.0 &  3.25 &  5.68 &  20.5 &  4.22 &  5.11 &  21.3 \\ \smallskip
 & $^{136}$Xe  &  7.83 &  2.11 &  3.70 &  13.3 &  2.74 &  3.33 &  13.9 \\ 
\multirow{5}{*}{AV-18}
 & $^{48}$Ca   &  2.28 &  0.43 &  0.82 &  2.40 &  0.58 &  1.00 &  8.21 \\ 
 & $^{76}$Ge   &  1.02 &  0.20 &  0.37 &  1.11 &  0.28 &  0.47 &  3.33 \\ 
 & $^{82}$Se   &  3.73 &  0.73 &  1.36 &  4.05 &  0.97 &  1.64 &  12.5 \\ 
 & $^{130}$Te  &  8.26 &  1.59 &  2.99 &  8.86 &  2.07 &  3.52 &  23.8 \\ 
 & $^{136}$Xe  &  5.41 &  1.04 &  1.95 &  5.75 &  1.34 &  2.30 &  15.5 \\  \hline \smallskip
&$T_{LNV}\cdot 10^{-26}$	&$T_{1}$&$T_{2}$	&$T_{3/RRz(LLz)}$&$T_{3/LRz(RLz)}$&$T_{4}$	&$T_{5}$&$T_{\pi N}$ \\ 
\multirow{4}{*}{CD-Bonn} 
 & $^{48}$Ca   &  2.53 &  2.54 &  2.54 &  2.54 &  2.46 &  2.45 &  2.01 \\ 
 & $^{76}$Ge   &  5.68 &  5.61 &  5.62 &  5.58 &  5.18 &  5.25 &  4.96 \\ 
 & $^{82}$Se   &  1.56 &  1.54 &  1.55 &  1.54 &  1.50 &  1.51 &  1.32 \\  \smallskip
 & $^{130}$Te  &  0.70 &  0.70 &  0.70 &  0.69 &  0.69 &  0.70 &  0.70 \\ 
\multirow{4}{*}{AV-18}
 & $^{48}$Ca   &  2.54 &  2.56 &  2.56 &  2.57 &  2.48 &  2.46 &  2.03 \\ 
 & $^{76}$Ge   &  5.67 &  5.57 &  5.59 &  5.52 &  5.14 &  5.24 &  4.99 \\ 
 & $^{82}$Se   &  1.55 &  1.53 &  1.53 &  1.52 &  1.48 &  1.50 &  1.33 \\ 
 & $^{130}$Te  &  0.70 &  0.70 &  0.70 &  0.69 &  0.69 &  0.70 &  0.70 \\   \hline
 \end{tabular}
 \end{table*}
 
 \subsection{Disentangling contributions to the $0\nu\beta\beta$ decay rate from half-life ratios}\label{ratios}
The analysis is based on choosing a pair of isotopes, calculating the ratio of half-lives for different mechanisms, and identifying results that stand out and do not overlap. This means that we use the figures to search for bars that have a noticeable gap between them and other higher or lower bars. Quite obviously, the lower values could also be important and one could inverse the ratio to better see the gap between them and other results.  
 
One can easily notice in the following figures and in the tables of the previous subsection that the ${\cal{M}}_{LNV}^2$ coefficients of $^{130}$Te are very close to those of $^{136}$Xe. This feature can be understood as due to the NME and PSF that enter these coefficients, which are very similar for both isotopes. The main consequence of this fact is that both half-lives are of the same order. Due to this resemblance, measuring any of these two nuclei is equally desirable as they can easily substitute each other in the analysis of half-lives. The downside of this feature is that the half-life ratio among themselves cannot provide us with information that would enable one to distinguish different contributions to the $0\nu\beta\beta$ rate. Although the predicted half-life for $^{130}$Te is slightly lower than that of $^{136}$Xe, and this could be favorable in some of the half-life ratios, the current experimental limits and trends lead us to believe that $^{136}$Xe and $^{76}$Ge are likely to be the first ones to be experimentally measured. 
Based on this assumption about the experimental expectations, we present our analysis in relation to these two isotopes. The reader can, however, use the calculated ${\cal{M}}_{LNV}^2$ coefficients listed in the tables of the previous Subsection to investigate the half-life ratios of any pairs of nuclei. 

\begin{figure*}  
    \centering
 \begin{minipage}{0.8\linewidth}
   \includegraphics[width=0.99\textwidth]{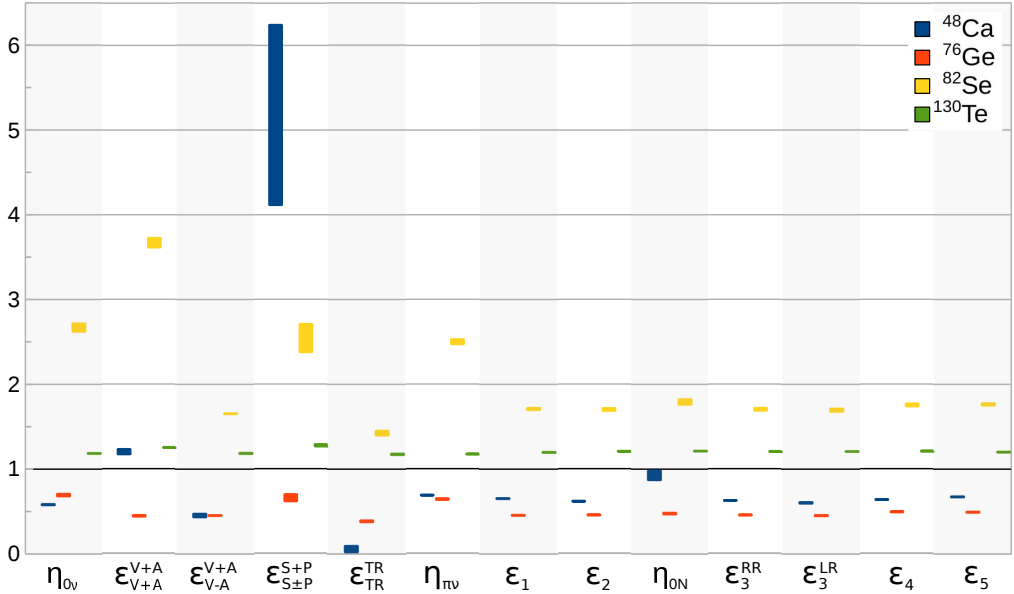}  
      \caption{The ratio between the $^{136}$Xe half-life and the $T_\alpha$ half-lives of several experimentally interesting isotopes, in the case of 12 EFT LNV couplings plus $\eta_{0N}$. The left to right order of the bars corresponds to up to down order in the Legend. The height of the bars represents the difference between results obtained with different SRC parameterizations. $\eta_{0N}$ plays a similar role to $\varepsilon_{3}^{RRz(LLz)}$.}
    \label{fig-xe-ratio-cmu} 
 \end{minipage}
\end{figure*}

In Fig.~\ref{fig-xe-ratio-cmu} we present the half-life ratio of $^{136}$Xe over another isotope for the mechanisms discussed, in the case of our preference of Hamiltonians (CMU). The bars connect two values for each ratio. One value is obtained using the CD-Bonn parametrization for the SRC, while the other value is the result of our calculations using the AV-18 parametrization. The graphical representation emphasizes the impact of the choice of SRC for the ratio of the half-lives. It is easy to notice that for most cases, the SRC plays an insignificant role for the half-life ratios, and the bars of the plots had to be increased for the reader to see them.  In the two cases where SRC did make a difference ($S+P/S\pm P$ and $TR/TR$ ), the bars are completely outside of the range of those provided by the other mechanisms, and the analysis is not affected by their spread. From this figure, it appears that the dominance of the $\epsilon^{S+P}_{S\pm P}$ and $\epsilon^{TR}_{TR}$ contributions could be confirmed or ruled out by the $^{136}$Xe/$^{48}$Ca ratio, while the $\epsilon^{V+A}_{V+A}$ could be investigated by the $^{136}$Xe/$^{82}$Se ratio. Then, the dominance of $\epsilon^{V+A}_{V-A}$ mechanism (know also as the $\eta$ mechanism) could be identified from the two-electron angular and energy distributions \cite{HoroiNeacsu2016prd,Neacsu2016ahep-dist}.

\begin{figure*}  
    \centering
 \begin{minipage}{0.8\linewidth}
   \includegraphics[width=0.99\textwidth]{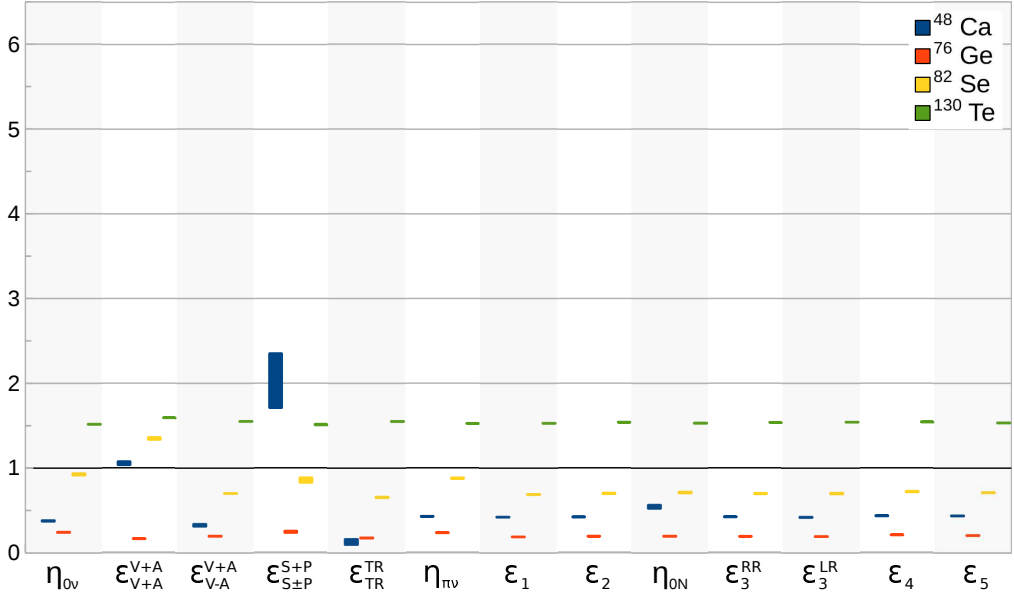}  
      \caption{Same as Fig.~\ref{fig-xe-ratio-cmu}, but for the Strasbourg-Madrid choice of Hamiltonians.}
    \label{fig-xe-ratio-stma} 
 \end{minipage}
\end{figure*}

The same analysis performed in Fig.~\ref{fig-xe-ratio-cmu} is done for the St-Ma choice of Hamiltonians, and we present those results in Fig.~\ref{fig-xe-ratio-stma}. The y-axis ranges are kept identical for an easier observation of the effect of changing the shell model Hamiltonians. Different from the previous figure is that the $^{130}$Te almost flat line has shifted higher, while the other ratios have decreased in magnitude. The $\epsilon^{S+P}_{S\pm P}$ and $\epsilon^{TR}_{TR}$ contributions can still be identifiable with the $^{136}$Xe/$^{48}$Ca ratio, but the $\epsilon^{V+A}_{V+A}$ from $^{136}$Xe/$^{82}$Se identification would be not as sensitive as for the CMU NME.

\begin{figure*}  
    \centering
 \begin{minipage}{0.8\linewidth}
   \includegraphics[width=0.99\textwidth]{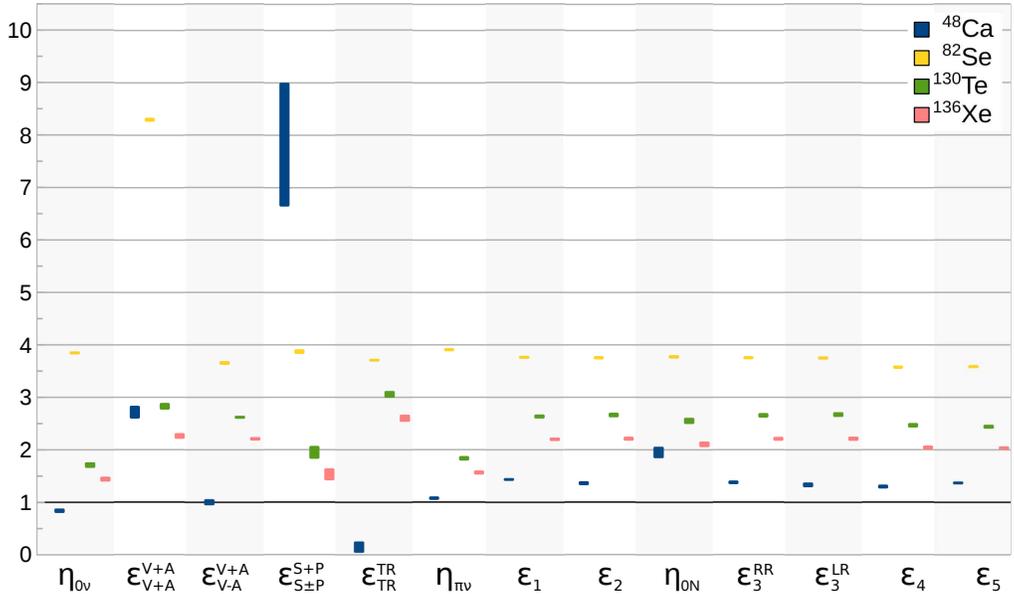}  
      \caption{Same as Fig.~\ref{fig-xe-ratio-cmu}, but for $^{76}$Ge instead of $^{136}$Xe, with the CMU choice of Hamiltonians.}
    \label{fig-ge-ratio-cmu} 
 \end{minipage}
\end{figure*}

Similar to Fig.~\ref{fig-xe-ratio-cmu}, we also represent in Fig.~\ref{fig-ge-ratio-cmu} the half-life ratios of $^{76}$Ge over those of other isotopes to search for potentially other identifiable mechanisms. As in the previous cases, it is easy to see the consistency of results using the same SRC. Because both $^{76}$Ge and $^{82}$Se can be calculated with the same Hamiltonian, we consider the ratios of half-lives for these nuclei to be the ones with the least uncertainties. Due to this feature, the dominance of the $\epsilon^{V+A}_{V+A}$ contribution could be reliably validated or ruled-out with this pair of isotopes. Very similar to the previous two figures, the $\epsilon^{S+P}_{S\pm P}$ and $\epsilon^{TR}_{TR}$ contributions could also be confirmed or ruled out by the $^{76}$Ge/$^{48}$Ca ratio.

\begin{figure*}  
    \centering
 \begin{minipage}{0.8\linewidth}
   \includegraphics[width=0.99\textwidth]{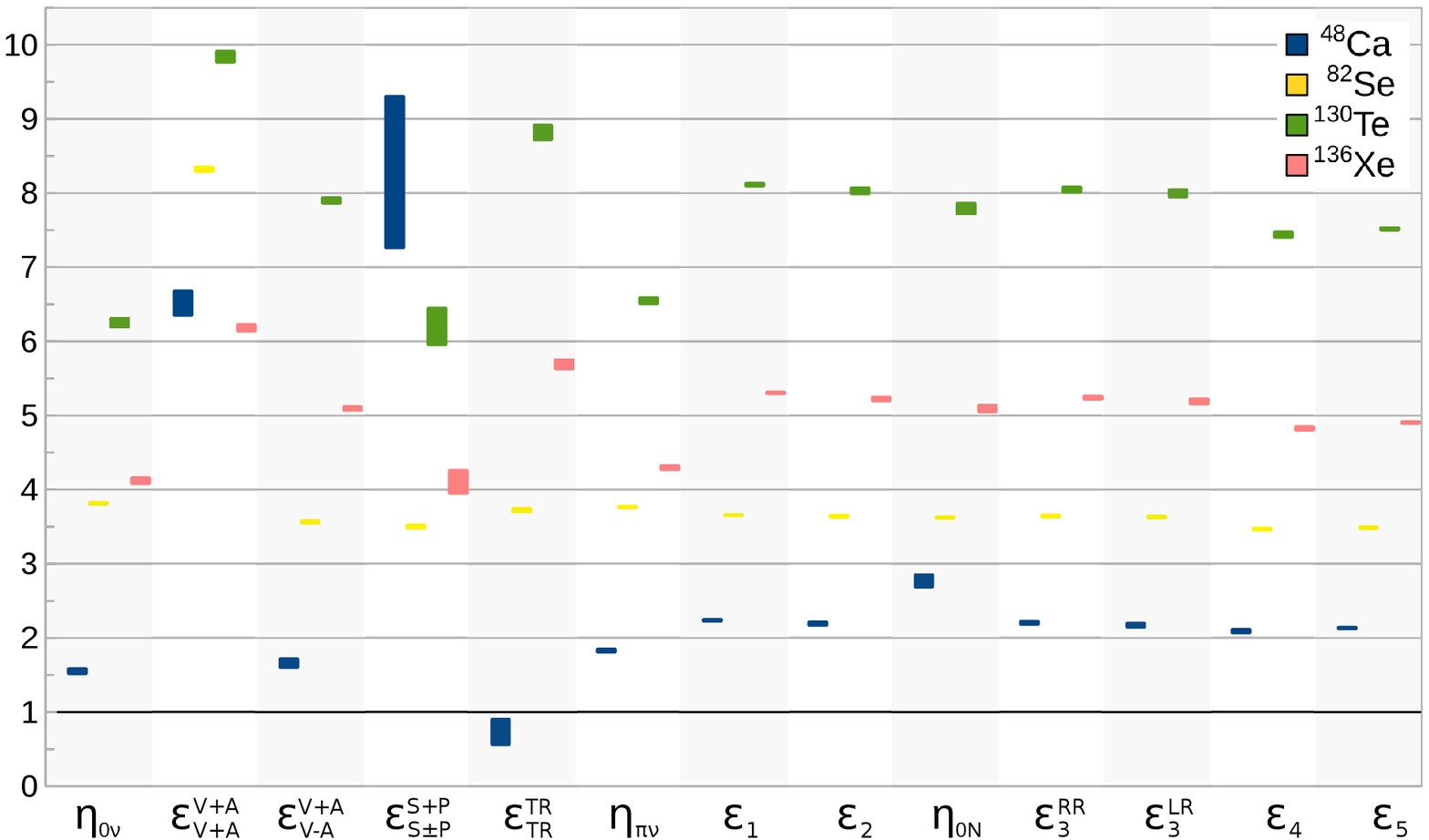}  
      \caption{Same as Fig.~\ref{fig-ge-ratio-stma}, but for the Strasbourg-Madrid effective Hamiltonians.}
    \label{fig-ge-ratio-stma} 
 \end{minipage}
\end{figure*}

As in Fig.~\ref{fig-ge-ratio-cmu}, we show the results for the St-Ma Hamiltonians in Fig.~\ref{fig-ge-ratio-stma} in the same y-axis range. From this image we could identify the $\epsilon^{V+A}_{V+A}$ contribution by the $^{76}$Ge/$^{82}$Se ratio, but the 
$\epsilon^{S+P}_{S\pm P}$ and $\epsilon^{TR}_{TR}$ mecanisms are more difficult to confim or rule out than in the previous figure using the $^{76}$Ge/$^{48}$Ca ratio.

In all cases presented, the short-range contributions corresponding to the dimension 9 Lagrangian cannot be disentangled from each other using ratios of half-lives. In the Tables and in the Figures the results for these cases correspond to $|\varepsilon_{\alpha}^{\beta}|$. None of the half-lives, or the ratios of half-lives, are different enough to be distinguishable from the others.  

\section{Conclusions} \label{conclusions}
If the neutrinoless double-beta decay process is experimentally confirmed, an analysis of possible contributions to the decay may be possible, based on the measured half-lives for different isotopes alone. The ratio of half-lives has been proposed before as a possible method for disentangling the heavy right-handed neutrino contribution from that of the light left-handed one. 

In this paper we performed an in-depth analysis of the ratio of half-lives for 5 isotopes that are actively considered by experimentalists. 
We took into account  12 contributions to the $0\nu\beta\beta$ described by the left-right symmetric model, the $R$-parity violating SUSY model, and an effective field theory. The study is presented under the assumption that only one contribution dominates.
One main conclusion is that the nuclear matrix elements need to be calculated with better accuracy for most of the ratios to be relevant. 

For a long time there were many debates in the literature regarding the uncertainties in the NME that originate  from the treatment of the short-range correlations (SRC). Several methods and parametrization were developed for their inclusion in $0\nu\beta\beta$ calculations. Depending on the choice of the SRC method and parameters, the changes to the NME ranged from about 20\% for the light left-handed neutrino exchange, to a dramatic ~50\% change in the case of the heavy right-handed neutrino exchange. In our investigation we found out that the SRC choice usually affects the half-life ratio only by around 1\%, with the exception of the "$S+P/S\pm P$" and "$TR/TR$" cases. Based on this observation, we conclude that the SRC do not significantly affect our analysis of the half-life ratios calculated with ISM NME, as long as the choice is consistent for all isotopes considered. This conclusion does not diminish the importance and the need for obtaining an effective transition operator that properly takes into account the SRC effects in a consistent manner, rather just rules-out most of the uncertainty related to SRC for this particular type of analysis.

What was shown to have a great impact on our study was the choice of effective shell model Hamiltonians. It not only changes the extracted LNV couplings, but also places a large uncertainty over several half-life ratios. The isotopes considered here are calculated in three different model spaces, using a total of six effective Hamiltonians. It is not always possible to have half-life ratios for isotopes calculated in the same model space and using the same Hamiltonian. In the two cases where this is possible, the pairs $^{76}\textmd{Ge} - ^{82}$Se and $^{130}\textmd{Te} - ^{136}$Xe, the ratios spread 
and the consistency of the calculations is quite high. Unfortunately, the ratio between $^{130}\textmd{Te}$ and $^{136}$Xe does not bring any information to this analysis, as it is constant, thus overlapping for all couplings.
However, when choosing isotopes from different model spaces, the half-life ratios for a pair of isotopes can spread significantly. One extreme case is that of $^{136}\textmd{Xe}/^{82}\textmd{Se}$ for the $\epsilon^{V+A}_{V+A}$ contribution where the ratio spreads from 1.3, when using GCN 28:50 for $^{82}\textmd{Se}$ and GCN 50:82 for $^{136}\textmd{Xe}$, to 3.6 in the case of JUN45 for $^{82}\textmd{Se}$ and SVD for $^{136}\textmd{Xe}$.

Nevertheless, with our present analysis one could, in principle, distinguish several of the contributions to the $0\nu\beta\beta$ process that stand out beyond the uncertainties that arise from using the shell model Hamiltonians discussed.
Ideally, more information could be extracted if the experimental half-lives of $^{48}$Ca, $^{76}$Ge, and $^{82}$Se become available. The half-life ratio $^{76}$Ge/$^{48}$Ca could indicate or rule-out the $\epsilon^{TR}_{TR}$ contribution and the $\epsilon^{S+P}_{S\pm P}$ contribution. Another possible contribution, that of $\epsilon^{V+A}_{V+A}$ (also corresponding to the $\eta_{\lambda}-$mechanism), could be investigated by the ratio of $^{76}$Ge/$^{82}$Se. If the tracking of the outgoing electrons will also become available, this complementary information could help decide the role of $\epsilon^{V+A}_{V-A}$ (also corresponding to the $\eta_{\eta}-$mechanism) \cite{HoroiNeacsu2016prd}. 

Based on the figures presented, we conclude that once better and more consistent NME calculations become available, and with the complementary information from electron angular and energy distributions, it could be possible to distinguish all the couplings in the dimension-6 Lagrangian if the half-lives of several isotopes are measured. The half-life ratio corresponding to the couplings in the dimension-9 Lagrangian would still remain inseparable from the ratio corresponding to heavy right-handed neutrino exchange ($\epsilon_{3}^{RR}$ in the EFT). Those need to be investigated via other methods, such as same charge dilepton production at LHC, etc.  
\\
 
\begin{acknowledgments}
Support from the U.S. Department of Energy Grants No. DE-SC0008529 and DE-SC0008641 is acknowledged. 
\end{acknowledgments}

\bibliographystyle{apsrev}
\bibliography{bb}

\end{document}